\documentclass{JHEP3}
\usepackage{graphics}
\newcommand{\be}{\begin{equation}}
\newcommand{\ee}{\end{equation}}
\newcommand{\bea}{\begin{eqnarray}}
\newcommand{\eea}{\end{eqnarray}}
\newcommand{\gm}{\gamma}
\newcommand{\Gm}{\Gamma}
\newcommand{\ep}{\varepsilon}
\newcommand{\lm}{\lambda}
\newcommand{\nn}{\nonumber}
\newcommand{\dd}{{\rm d}}
\newcommand{\Li}[1]{\mathop{\mathrm{Li}}\nolimits_{#1}}
\newlength{\prepwidth}

\title{Decoupling of heavy quarks in HQET}

\author{Andrey G.~Grozin\\
Institut f\"ur Theoretische Teilchenphysik\\
Universit\"at Karlsruhe, Germany\\
E-mail: \email{grozin@particle.uni-karlsruhe.de}}

\author{Alexander V.~Smirnov\\
Scientific Research Computing Center of Moscow State University\\
Moscow 119992, Russia\\
E-mail: \email{asmirnov@rdm.ru}}

\author{Vladimir A.~Smirnov\\
Nuclear Physics Institute of Moscow State University\\
Moscow 119992, Russia\\
E-mail: \email{smirnov@theory.sinp.msu.ru}}

\abstract{Decoupling of $c$-quark loops in $b$-quark HQET is considered.
The decoupling coefficients for the HQET heavy-quark field
and the heavy--light quark current are calculated with the three-loop accuracy.
The last result can be used to improve the accuracy
of extracting $f_B$ from HQET lattice simulations
(without $c$-quark loops).
The decoupling coefficient for the flavour-nonsinglet QCD current
with $n$ antisymmetrized $\gamma$-matrices is also obtained at three loops;
the result for the tensor current ($n=2$) is new.}

\preprint{\settowidth{\prepwidth}{SFB/CPP-06-46}%
\begin{minipage}{\prepwidth}
TTP06-25\\
SFB/CPP-06-46\\
hep-ph/0609280
\end{minipage}}

\begin{document}

\newpage

\section{Introduction}
\label{S:Intro}

Let us consider QCD with a heavy flavour, say $c$.
It is well-known that processes with light quarks and gluons
having characteristic momenta much less than $m_c$
can be described by an effective low-energy theory ---
QCD without $c$-quarks.
The renormalized light-quark and gluon fields $q_i(\mu)$, $A(\mu)$
in the full theory
are related to the corresponding fields in the effective theory,
up to corrections suppressed by powers of $1/m_c$,
by the decoupling relations
\begin{equation}
q_i(\mu) = \zeta_q^{1/2}(\mu) q_i'(\mu)\,,
\qquad
A(\mu) = \zeta_A^{1/2}(\mu) A'(\mu)\,,
\label{Intro:Decq}
\end{equation}
where all quantities in the low-energy theory are denoted by primes
(we use the $\overline{\mathrm{MS}}$ renormalization scheme
throughout this paper).
Similarly, the coupling constant, the gauge-fixing parameter
and the light-quark masses in the two theories are related by
\begin{equation}
\alpha_s(\mu) = \zeta_\alpha(\mu) \alpha_s'(\mu)\,,
\qquad
a(\mu) = \zeta_A(\mu) a'(\mu)\,,
\qquad
m_i(\mu) = \zeta_m(\mu) m_i'(\mu)\,.
\label{Intro:DecL}
\end{equation}
The QCD decoupling coefficients are known at three~\cite{CKS:98}
and even four loops~\cite{CKS:05,SS:06a}%
\footnote{The result of~\cite{SS:06a} contains one master integral
which was not known analytically, only numerically, with 37-digits precision.
An analytical expression for this integral
has been published later~\cite{KKOV:06}.}.
Various operators of full QCD, e.g., light--light quark currents,
can be expressed via operators of the low-energy effective theory,
similarly to~(\ref{Intro:Decq}).

Now let us consider the $b$-quark HQET.
If the characteristic residual momentum of $b$,
as well as characteristic momenta of light quarks and gluons,
are much less than $m_c$, then all processes can be described
by the low-energy effective theory --- HQET without $c$-quarks.
The decoupling relations for the light fields~(\ref{Intro:Decq})
and the parameters of the Lagrangian~(\ref{Intro:DecL})
are exactly the same as in QCD.
The static $b$-quark field $\tilde{Q}$
in the ``full'' theory (HQET with $c$-quarks)
and in the effective theory (HQET without $c$-quarks)
are related by
\begin{equation}
\tilde{Q}(\mu) = \tilde{\zeta}_Q^{1/2}(\mu) \tilde{Q}'(\mu)\,.
\label{Intro:DecQ}
\end{equation}
Various operators of the ``full'' HQET can be expressed
via operators of the low-energy HQET.
For example, for the heavy--light quark currents
$\tilde{\jmath}=\bar{q}\Gamma\tilde{Q}$ we have
\begin{equation}
\tilde{\jmath}(\mu) = \tilde{\zeta}_j(\mu) \tilde{\jmath}'(\mu)\,,
\label{Intro:Decj}
\end{equation}
up to corrections suppressed by powers of $1/m_c$.
The HQET-specific decoupling coefficients
$\tilde{\zeta}_Q$, $\tilde{\zeta}_j$
have been calculated in~\cite{G:98}
with the two-loop accuracy%
\footnote{One more HQET decoupling coefficient,
that for the $b$-quark chromomagnetic interaction,
has been found in~\cite{G:00a},
based on the calculations of~\cite{CG:97},
also at two loops.}.
Decoupling in HQET is also discussed in~\cite{G:04},
Sects.~4.7 and~5.5.

In the present paper, we shall calculate
$\tilde{\zeta}_Q$ and $\tilde{\zeta}_j$
up to three loops.
To do so, we need to calculate
on-shell HQET propagator diagrams
containing a massive quark loop.
Reduction of scalar Feynman integrals of this type
to master integrals is considered in Sect.~\ref{S:Red};
the master integrals are calculated in Sect.~\ref{S:Master}.
The HQET decoupling coefficients are obtained
in Sects.~\ref{S:DQ} and~\ref{S:Dj}.
In Appendix~\ref{S:QCD} we derive the three-loop decoupling coefficient
for the QCD flavour-nonsinglet quark current with $n$ antisymmetrized $\gamma$-matrices,
as a generic formula containing $n$;
the result for the tensor current ($n=2$) is new.

\newpage

\section{Reduction of Feynman integrals}
\label{S:Red}

\subsection{General remarks}
\label{S:OS}

\FIGURE{
\begin{picture}(52,26)
\put(26,14.5){\makebox(0,0){\includegraphics{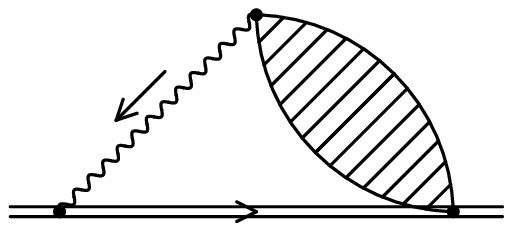}}}
\put(26,3){\makebox(0,0)[t]{$k$}}
\put(11,19){\makebox(0,0){$k$}}
\put(25,7){\makebox(0,0)[b]{1}}
\put(17.5,12.5){\makebox(0,0){2}}
%\put(36,15){\makebox(0,0){$f(k^2)$}}
\end{picture}
\caption{Diagram with a single HQET line}
\label{F:aver}}

In this subsection, we discuss some general properties
and simple particular cases of on-shell HQET integrals
with a massive quark loop.

Let us consider the integrals (Fig.~\ref{F:aver})
\begin{eqnarray}
&&F(a_1,a_2) = \int \frac{f(k^2)\,d^d k}{E_1^{a_1} E_2^{a_2}}\,,
\nonumber\\
&&E_1 = -2k\cdot v-i0\,,\qquad
E_2 = -k^2-i0
\label{OS:Ff}
\end{eqnarray}
with some $f(k^2)$.
Following the strategy of integration by parts~\cite{CT:81}
and integrating the identity
\begin{equation}
\frac{\partial}{\partial k} \cdot
\left( k - 2 \frac{E_2}{E_1} v \right)
\frac{f(k^2)}{E_1^{a_1} E_2^{a_2}}
= \left[ d-a_1-2 - 4 (a_1+1) \frac{E_2}{E_1^2} \right]
\frac{f(k^2)}{E_1^{a_1} E_2^{a_2}}\,,
\label{OS:idd}
\end{equation}
we obtain the recurrence relation
\begin{equation}
(d-a_1-2) F(a_1,a_2) =
4 (a_1+1) \mathbf{1}^{++} \mathbf{2}^- F(a_1,a_2)
\label{OS:idr}
\end{equation}
(here, as usual, the operator $\mathbf{2}^-$ decreases $a_2$ by 1,
and $\mathbf{1}^{++}$ increases $a_1$ by 2).
Its solution is
\begin{equation}
F(a_1,a_2) =
\left\{
\begin{array}{ll}
\displaystyle (-4)^{-a_1/2}
\frac{\Gamma\left(\frac{d}{2}\right)}{\Gamma\bigl(\frac{d-a_1}{2}\bigr)}
\frac{\Gamma\left(\frac{1-a_1}{2}\right)}{\Gamma\left(\frac{1}{2}\right)}
F\Bigl(0,a_2+\frac{a_1}{2}\Bigr)\,,
& \mbox{even $a_1$,}\\
\displaystyle 2^{1-a_1}
\frac{\Gamma\left(\frac{d-1}{2}\right)}%
{\Gamma\left(\frac{a_1+1}{2}\right)\Gamma\bigl(\frac{d-a_1}{2}\bigr)}
F\Bigl(1,a_2+\frac{a_1-1}{2}\Bigr)\,,
& \mbox{odd $a_1>0$,}\\
\displaystyle 0, & \mbox{odd $a_1<0$,}
\end{array}
\right.
\label{OS:aver}
\end{equation}
The result for even $a_1\le0$ can easily be derived
by averaging over $v$ directions in the $d$-dimensional
Euclidean space after the Wick rotation;
it was supposed in~\cite{BG:95} that the same formula
can be used for even $a_1>0$,
and this method was used, e.g., in~\cite{CG:97,G:98},
but no proof existed.
Now we see that this averaging trick follows from~(\ref{OS:idr}).
For odd $a_1<0$, averaging over $v$ directions gives $0$,
but this result does not extend to odd $a_1>0$.

\FIGURE{
\begin{picture}(52,26)
\put(26,14.5){\makebox(0,0){\includegraphics{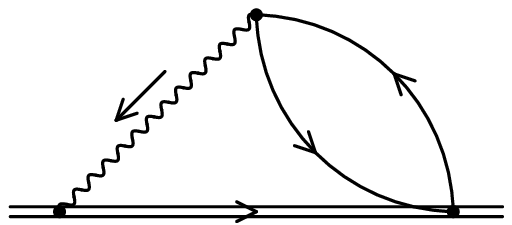}}}
\put(26,3){\makebox(0,0)[t]{$k_1$}}
\put(11,19){\makebox(0,0){$k_1$}}
\put(29.5,9.5){\makebox(0,0){$k_2$}}
\put(49,20){\makebox(0,0){$k_1+k_2$}}
\put(25,7){\makebox(0,0)[b]{1}}
\put(17.5,12.5){\makebox(0,0){2}}
\put(34,13){\makebox(0,0){3}}
\put(38,17){\makebox(0,0){4}}
\end{picture}
\caption{Two-loop diagram}
\label{F:l2}}

Now we shall discuss the two-loop integrals (Fig.~\ref{F:l2})
\begin{equation}
F(a_1,a_2,a_3,a_4,a_5) = \frac{1}{(i\pi^{d/2})^2} \int
\frac{d^d k_1\,d^d k_2}{E_1^{a_1} E_2^{a_2} E_3^{a_3} E_4^{a_4} E_5^{a_5}}
\label{OS:l2}
\end{equation}
with four denominators
\begin{eqnarray}
&&E_1 = -2 k_1\cdot v - i0\,,\qquad
E_2 = -k_1^2 - i0\,,
\nonumber\\
&&E_3 = 1 - k_2^2 - i0\,,
\nonumber\\
&&E_4 = 1 - (k_1+k_2)^2 - i0
\label{OS:E14}
\end{eqnarray}
(we have put the quark mass $m=1$;
the power of $m$ can easily be restored by dimensionality),
and one numerator
\begin{equation}
E_5 = (2 k_2 + k_1)\cdot v
\label{OS:E5}
\end{equation}
($a_5$ is always $\le0$).
Integrals with $a_3\le0$ or $a_4\le0$ vanish.
We have
\begin{equation}
F(a_1,a_2,a_3,a_4,a_5) = (-1)^{a_5} F(a_1,a_2,a_4,a_3,a_5)\,.
\label{OS:sym}
\end{equation}

These integrals can be subdivided into two disjoint subsets,
with even and odd $a_1+a_5$.
We shall call them ``apparently even'' and ``apparently odd''
(they would be even and odd with respect to $v\to-v$
if there were no $-i0$ in $E_1$).
Apparently odd integrals with $a_1\le0$ vanish,
because we can omit $-i0$ in the numerator.
These two classes are not mixed by any recurrence relations,
therefore, we have two disjoint problems.

These integrals can be calculated as follows.
First, we get rid of the numerator $E_5^{-a_5}$.
We have a contraction of $-a_5$ vectors $v$
with the one-loop tensor integral which depends on $k_1$.
Decomposing this integral into all possible tensor structures,
we obtain an expression containing only $E_1$ (no $E_5$).
Using $\alpha$-parametrization, we obtain
\begin{eqnarray}
&&F(a_1,a_2,a_3,a_4,0) = {}
\label{OS:res}\\
&&\frac{\Gamma\left(\frac{a_1}{2}\right)
\Gamma\left(\frac{a_1-d}{2}+a_2+a_3\right)
\Gamma\left(\frac{a_1-d}{2}+a_2+a_4\right)
\Gamma\Bigl(\frac{a_1}{2}+a_2+a_3+a_4-d\Bigr)
\Gamma\left(\frac{d-a_1}{2}-a_2\right)}%
{2\Gamma(a_1)\Gamma(a_3)\Gamma(a_4)
\Gamma(a_1+2a_2+a_3+a_4-d)
\Gamma\left(\frac{d-a_1}{2}\right)}\,.
\nonumber
\end{eqnarray}
In full accordance with~(\ref{OS:aver}),
integrals $F(a_1,a_2,a_3,a_4,0)$ with even $a_1$
reduce to $F(0,a_2+a_1/2,a_3,a_4,0)$
(this is a well-known two-loop vacuum integral~\cite{V:80});
those with odd $a_1>0$ reduce to $F(1,a_2+(a_1-1)/2,a_3,a_4,0)$;
and those with odd $a_1<0$ vanish.
This is a strong check.
All apparently even integrals are proportional to the single master integral
\begin{equation}
I_0^2 = \raisebox{-9.8mm}{\includegraphics{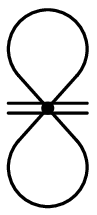}}\,,
\label{OS:I02}
\end{equation}
where
\begin{equation}
I_0 = \frac{1}{i\pi^{d/2}} \int
\frac{d^d k}{1-k^2-i0}
= \Gamma(1-d/2)
\label{OS:I0}
\end{equation}
is the one-loop vacuum integral.
All apparently odd integrals are proportional to the single master integral
\begin{equation}
J_0 = \raisebox{-5.8mm}{\includegraphics{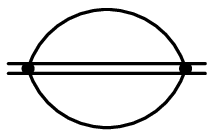}}
= 2^{4d-9} \pi^2 \frac{\Gamma(5-2d)}{\Gamma^2(2-d/2)}\,.
\label{OS:J0}
\end{equation}

In order to perform our decoupling calculation,
we need two classes of three-loop integrals.
They will be reduced to master integrals
using integration by parts~\cite{CT:81}.
A similar problem (off-shell HQET integrals
without massive quarks)
has been solved earlier~\cite{G:00}.
However, constructing reduction algorithms by hand
is tedious and time-consuming.
Several algorithmic methods to solve reduction problems
for Feynman integrals have been recently suggested;
they are discussed, e.g., in~\cite{S4}.
Some approaches are based on the use of Gr\"obner
bases~\cite{Tar,SS:06,S:06}.
This is the last algorithm in this list~\cite{SS:06,S:06}
that we have used in the present calculations.

\subsection{Integrals of class 1}
\label{S:1}

\FIGURE{
\begin{picture}(92,45)
\put(46,22.5){\makebox(0,0){\includegraphics{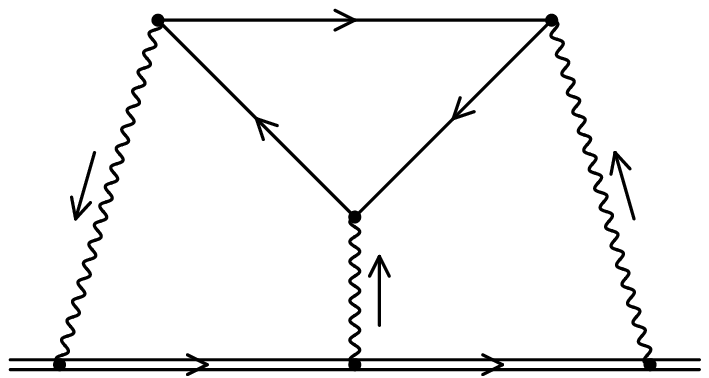}}}
\put(31,3){\makebox(0,0)[t]{$k_1$}}
\put(61,3){\makebox(0,0)[t]{$k_2$}}
\put(46,42){\makebox(0,0)[b]{$k_3$}}
\put(50,12.5){\makebox(0,0)[l]{$k_1-k_2$}}
\put(17,24){\makebox(0,0)[r]{$k_1$}}
\put(75,24){\makebox(0,0)[l]{$k_2$}}
\put(37,27){\makebox(0,0)[r]{$k_1+k_3$}}
\put(55,27){\makebox(0,0)[l]{$k_2+k_3$}}
\put(30,7){\makebox(0,0)[b]{1}}
\put(60,7){\makebox(0,0)[b]{2}}
\put(22.5,21){\makebox(0,0)[l]{3}}
\put(69.5,21){\makebox(0,0)[r]{4}}
\put(44,12.5){\makebox(0,0)[r]{5}}
\put(46,38){\makebox(0,0)[t]{6}}
\put(38.5,31.5){\makebox(0,0)[l]{7}}
\put(53.5,31.5){\makebox(0,0)[r]{8}}
\end{picture}
\caption{Diagram of class 1}
\label{F:1}}

These integrals (Fig.~\ref{F:1}),
\begin{equation}
F_1(a_1,a_2,a_3,a_4,a_5,a_6,a_7,a_8,a_9) = \frac{1}{(i\pi^{d/2})^3} \int
\frac{d^d k_1\,d^d k_2\,d^d k_3}{\prod_{i=1}^9 E_i^{a_i}}\,,
\label{1:def}
\end{equation}
have 8 denominators,
\begin{eqnarray}
&&E_1 = - 2 k_1\cdot v\,,\qquad
E_2 =  - 2 k_2\cdot v\,,
\nonumber\\
&&E_3 = - k_1^2\,,\qquad
E_4 = - k_2^2\,,\qquad
E_5 = - (k_1-k_2)^2\,,
\nonumber\\
&&E_6 = 1 - k_3^2\,,\qquad
E_7 = 1 - (k_1+k_3)^2\,,\qquad
E_8 = 1 - (k_2+k_3)^2
\label{1:den}
\end{eqnarray}
(here and in what follows, $-i0$ is implied in all denominators),
and one numerator,
\begin{equation}
E_9 = 2 k_3\cdot v
\label{1:num}
\end{equation}
($a_9$ is always $\le0$).
The integral~(\ref{1:def}) is symmetric with respect to
$(1\leftrightarrow2,3\leftrightarrow4,7\leftrightarrow8)$.
It vanishes if indices of the following subsets of lines
are non-positive:
$\{5,7\}$, $\{5,8\}$, $\{6,7\}$, $\{6,8\}$, $\{7,8\}$,
$\{3,4,6\}$.

We applied the method of~\cite{SS:06,S:06} to construct
an algorithm for reducing integrals of this class
to master integrals.
Recurrence relations do not mix apparently even integrals
(with even $a_1+a_2+a_9$) with apparently odd ones.
All apparently even integrals reduce to linear combinations
of the following master integrals:
\begin{eqnarray}
&&I_0^3 = \raisebox{-6.3mm}{\includegraphics{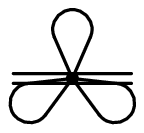}}\,,\qquad
I_1 = \raisebox{-6.3mm}{\includegraphics{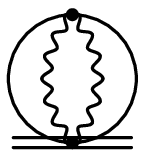}}\,,\qquad
I_2 = \raisebox{-4.3mm}{\includegraphics{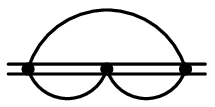}}\,,
\nonumber\\
&&I_3 = \raisebox{-3.8mm}{\includegraphics{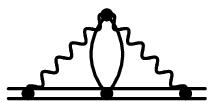}}\,,\qquad
I_4 = \raisebox{-3.8mm}{\includegraphics{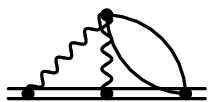}}\,.
\label{1:even}
\end{eqnarray}
% The program constructing Gr\"obner-like bases for this problem
% considered two more integrals as master.
% But they are proportional to $I_1$ with rational coefficients.
% The first one is
% \begin{equation}
% \bar{J}_1 = \raisebox{-5.8mm}{\includegraphics{jb1.eps}}
% = \frac{d-2}{4} I_1\,,
% \label{1:Jb1}
% \end{equation}
% using~(\ref{OS:aver}).
% The second one is
% \begin{equation}
% \bar{J}_2 = F_1(-1,1,1,0,1,1,0,1,0) =
% \raisebox{-3.8mm}{\includegraphics{j2.eps}} (-2 k_1\cdot v)
% \label{1:Jb2}
% \end{equation}
% (we mean that the numerator $-2 k_1\cdot v$ is under the integral sign).
% Here $v$ is contracted with a vector one-loop massless integral
% which depends only on $k_2$, and hence is directed along $k_2$.
% We may replace $-2k_1\cdot v\to-2k_2\cdot v\,k_1\cdot k_2/k_2^2$,
% and the HQET denominator $E_2$ cancels.
% Re-writing $k_1\cdot k_2$ as $[k_1^2+k_2^2-(k_1-k_2)^2)/2$,
% we obtain two vanishing integrals plus $I_1/2$:
% \begin{equation}
% \bar{J}_2 = \frac{1}{2} I_1\,.
% \label{1:Jb2res}
% \end{equation}
All apparently odd integrals reduce to the following master ones:
\begin{equation}
I_0 J_0 = \raisebox{-6.3mm}{\includegraphics{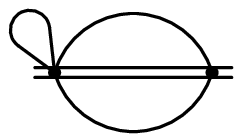}}\,,\qquad
J_1 = \raisebox{-5.8mm}{\includegraphics{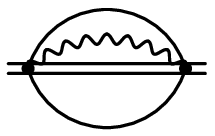}}\,,\qquad
J_2 = \raisebox{-3.8mm}{\includegraphics{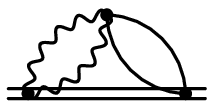}}\,.
\label{1:odd}
\end{equation}

\subsection{Integrals of class 2}
\label{S:2}

\FIGURE{
\begin{picture}(96,36)
\put(46,20){\makebox(0,0){\includegraphics{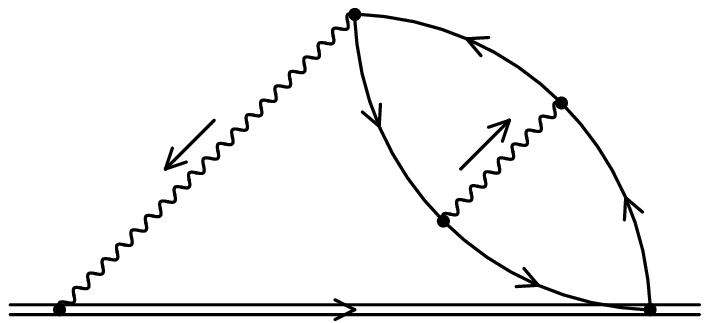}}}
\put(46,3){\makebox(0,0)[t]{$k_3$}}
\put(26,24){\makebox(0,0){$k_3$}}
\put(45,24){\makebox(0,0){$k_1$}}
\put(59,8){\makebox(0,0){$k_2$}}
\put(64,35){\makebox(0,0){$k_1+k_3$}}
\put(82,17){\makebox(0,0){$k_2+k_3$}}
\put(56,27){\makebox(0,0){$k_1-k_2$}}
\put(45,7){\makebox(0,0)[b]{1}}
\put(32.5,17.5){\makebox(0,0){2}}
\put(52,22){\makebox(0,0){3}}
\put(66,10){\makebox(0,0){4}}
\put(56,30){\makebox(0,0){5}}
\put(71,13){\makebox(0,0){6}}
\put(63,18){\makebox(0,0){7}}
\end{picture}
\caption{Diagram of class 2}
\label{F:2}}

These integrals (Fig.~\ref{F:2}),
\begin{equation}
F_2(a_1,a_2,a_3,a_4,a_5,a_6,a_7,a_8,a_9)
= \frac{1}{(i\pi^{d/2})^3} \int
\frac{d^d k_1\,d^d k_2\,d^d k_3}{\prod_{i=1}^9 E_i^{a_i}}\,,
\label{2:def}
\end{equation}
have 7 denominators,
\begin{eqnarray}
&&E_1 = - 2 k_3\cdot v\,,\qquad
E_2 =  - k_3^2\,,\qquad
E_3 = 1 - k_1^2\,,\qquad
E_4 = 1 - k_2^2\,,
\nonumber\\
&&E_5 = 1 - (k_1+k_3)^2\,,\qquad
E_6 = 1 - (k_2+k_3)^2\,,\qquad
E_7 = - (k_1-k_2)^2\,,
\label{2:den}
\end{eqnarray}
and two numerators,
\begin{equation}
E_8 = (2k_1+k_3)\cdot v\,,\qquad
E_9 = (2k_2+k_3)\cdot v
\label{2:num}
\end{equation}
($a_8$ and $a_9$ are always $\le0$).
The integral~(\ref{2:def}) is symmetric with respect to
$(3\leftrightarrow4,5\leftrightarrow6,8\leftrightarrow9)$, and
\begin{equation}
F_2(a_1,a_2,a_3,a_4,a_5,a_6,a_7,a_8,a_9) =
(-1)^{a_8+a_9} F_2(a_1,a_2,a_5,a_6,a_3,a_4,a_7,a_8,a_9)\,.
\label{2:sym}
\end{equation}
It vanishes if indices of the following subsets of lines
are non-positive:
$\{3,5\}$, $\{4,6\}$, $\{3,4\}$, $\{5,6\}$.

We have used the method of~\cite{SS:06,S:06} to construct
an algorithm reducing these integrals to master ones.
For apparently even integrals (with even $a_1+a_8+a_9$)
we also used a method similar to subsection~\ref{S:OS}.
Namely, we have a contraction of $-a_8-a_9$ vectors $v$
with a two-loop tensor integral which depends on $k_3$.
Decomposing this integral into all possible tensor structures,
we get rid of the numerator $E_8^{-a_8}E_9^{-a_9}$.
Then we use the averaging formula~(\ref{OS:aver}),
and the problem reduces to three-loop massive vacuum integrals
$B_M$~\cite{B:92}.
A REDUCE package RECURSOR~\cite{B:92} reduces them
to two master integrals, $I_0^3$ (see~(\ref{1:even})) and
\begin{equation}
I_5 = \raisebox{-6.3mm}{\includegraphics{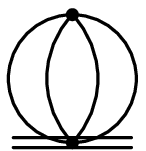}}\,.
\label{2:even}
\end{equation}
Agreement of results produced by these two ways
serves as a strong check.

Apparently odd integrals of this class are expressed
via $I_0 J_0$, $J_1$ (see~(\ref{1:odd})), and
\begin{equation}
J_3 = \raisebox{-3.8mm}{\includegraphics{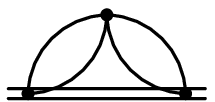}}\,.
\label{2:odd}
\end{equation}

\section{Master integrals}
\label{S:Master}

\subsection{Simple master integrals}
\label{S:Simple}

Simplest master integrals can exactly be expressed
via $\Gamma$ functions:
\begin{eqnarray}
I_1 &=& -
\frac{\Gamma(d/2-1)\Gamma(1-d/2)\Gamma^2(3-d)\Gamma(4-3d/2)}%
{\Gamma(6-2d)}\,,
\nonumber\\
J_1 &=& 2^{8d-17} \pi^2
\frac{\Gamma(d/2-1)\Gamma(5-2d)\Gamma(7-3d)}%
{\Gamma(2-d/2)\Gamma(3-d)\Gamma(4-3d/2)}\,,
\nonumber\\
J_2 &=& \frac{\pi^{1/2} \Gm^2(7/2 - d) \Gm(2 - d/2) \Gm^2(d/2-1)
  \Gm((3 (3 - d))/2) \Gm(d-5/2)}{2 \Gm(7 - 2 d) \Gm(d-2)
  \Gm((d-1)/2)}\,.
\label{Simple:Gamma}
\end{eqnarray}

The vacuum integral $I_5$~(\ref{2:even})
has been investigated in~\cite{B:92}.
It can be written as
\begin{eqnarray}
I_5 &=& - 4 \frac{(d-2)^2 (3d-11)}{(d-3)(3d-8)(3d-10)} I_0^3
- 7 \frac{d-4}{2d-7} I_1
\nonumber\\
&&{} + 64 \frac{(d-4)^2}{(d-2)(d-3)(3d-8)(3d-10)}
\Gamma^3(1+\varepsilon) B_4(\varepsilon)\,,
\label{Simple:I5}
\end{eqnarray}
where $d=4-2\varepsilon$, $B_4(\varepsilon)$ can be expressed via ${}_3F_2$ of unit
argument,
and
\begin{eqnarray}
B_4(\varepsilon) &=& B_4 + \mathcal{O}(\varepsilon)\,,
\nonumber\\
B_4 &=& 16 \Li4\left(\frac{1}{2}\right)
+ \frac{2}{3} \ln^2 2 \left( \ln^2 2 - \pi^2 \right)
- \frac{13}{180} \pi^4\,;
\label{Simple:B4}
\end{eqnarray}
its expansion up to the $\varepsilon^3$ term
was presented in~\cite{B:96}.

\subsection{Integral $I_2$}
\label{S:I2}

We evaluated all the complicated master integrals (i.e. which are
not expressed in terms of gamma functions for general $d$) by the
method of MB representation which is based on the following
formula
\be
\frac{1}{(X+Y)^{\lm}} = \frac{1}{\Gm(\lm)}
\frac{1}{2\pi i}\int_{-i \infty}^{+i \infty} \dd z\,
\Gm(\lm+z) \Gm(-z) \frac{Y^z}{X^{\lm+z}}  \; .
\label{MB}
\ee

Here the contour of integration is chosen in the standard way:
the poles with a $\Gm(\ldots+z)$ dependence
(let us call them {\em left} poles, for brevity)
are to the left of the contour and
the poles with a $\Gm(\ldots-z)$ dependence
({\em right} poles) are
to the right of it.
This formula is used to replace a sum of terms raised to some
power by their products raised to some powers at the cost of
having an extra integration.

Two systematic strategies for evaluating MB integrals in a Laurent
expansion in $\ep$ were suggested in~\cite{MB1,MB2}.
A description of this method is presented in~\cite{S:02}
and Chap.~4 of~\cite{S4}.

The simplest variant of using (\ref{MB}) is to write
down a massive propagator in terms of massless
ones~\cite{BD:91}:
\be
\frac{1}{(m^2-k^2)^\lm} = \frac{1}{\Gm(\lm)}
\frac{1}{2\pi i}\int_{-i \infty}^{+i \infty} \dd z
\frac{(m^2)^z}{(-k^2)^{\lm+z}} \Gm(\lm+z) \Gm(-z) \; .
\label{MBprop}
\ee

To have more checks, it is very
useful to derive such MB representations for general indices. To
evaluate $I_2$ we derived a MB representation for
$F_1(a_1,a_2,0,0,0,a_6,a_7,a_8,0)$.
We applied~(\ref{MBprop}) to the last two factors in the integrand
(with $a_7$ and $a_8$) and
evaluated the two resulting integrals over $k_1$ and $k_2$ and the
resulting integral over $k_3$ by
well-known one loop integration formulae to obtain
\bea
F_1(a_1,a_2,0,0,0,a_6,a_7,a_8,0)
%F(\lm_1,\ldots,\lm_5)
=\frac{1%\left(i\pi^{d/2} \right)^3
}{4 \sqrt{\pi} %(m^2)^{a_{12}/2+a_{678}+3\ep-6}(v^2)^{a_{12}/2}
\prod \Gm(a_i)}
\nn \\ && \hspace*{-80mm}
\times
\frac{1}{(2\pi i)^2} \int_{-i\infty}^{+i\infty} \int_{-i\infty}^{+i\infty}
\dd z_1\dd z_2\;\Gm(a_{12}/2 + a_{678} + 3 \ep -6+ z_1 + z_2)
\nn \\ &&  \hspace*{-80mm}\times
\frac{\Gm(a_1/2 + a_7 + \ep-2 + z_1)\Gm(a_2/2 + a_8 + \ep-2 + z_2)
 }
{\Gm((a_1 + a_2-7)/2 + a_7 + a_8 + 2 \ep + z_1 + z_2)}
\nn \\ &&  \hspace*{-80mm}\times
\Gm((a_1-3)/2 + a_7 + \ep + z_1)\Gm((a_2-3)/2 + a_8 + \ep + z_2)
\nn \\ &&  \hspace*{-80mm}\times
\Gm(2 - a_7 - \ep - z_1) \Gm(2 - a_8 - \ep - z_2)\Gm(-z_1)\Gm(-z_2)
\;,
\label{prob4-03-eq1}
\eea
where $a_{12}=a_1+a_2$ etc.

It was more convenient to obtain a Laurent expansion of $F_1(1,1,0,0,0,1,1,1,0)$ in $\ep$
from a Laurent expansion of the integral $F_1(2,2,0,0,0,1,2,2,0)$
and its reduction, in particular, to $I_2$.
For this auxiliary integral, we used (\ref{prob4-03-eq1}).
The resulting
twofold MB integral can be expanded immediately in $\ep$ because
there is  no gluing of poles of different nature when $\ep \to 0$.
Then one can close the integration contours over $z_1$ and $z_2$
to the right and obtain a double series. Its summation gives
the following result:
\bea
F_1(2,2,0,0,0,1,2,2,0) = \Gamma^3(1+\varepsilon)
\left(\frac{\pi^2}{18}-\frac{1}{3}
-\frac{\pi^2}{9}\ep +\mathcal{O}(\ep^2) \right)
\label{prob4-03-eq2}
\eea
which leads to
\bea
I_2 = - \Gamma^3(1+\varepsilon)\,\frac{\pi^2}{6}
\left(
\frac{1}{\ep}+\frac{5}{2} +\mathcal{O}(\ep) \right)\;.
\label{I2res}
\eea

Integrals $F_1(a_1,a_2,0,0,0,a_6,a_7,a_8,0)$ can be calculated
also in the coordinate space.
After continuation to Euclidean time, we have
\begin{eqnarray}
&&F_1(a_1,a_2,0,0,0,a_6,a_7,a_8,0) =
\frac{2^{3d/2-a_1-a_2-a_6-a_7-a_8+3}}%
{\Gamma(a_1)\Gamma(a_2)\Gamma(a_6)\Gamma(a_7)\Gamma(a_8)}
\nonumber\\
&&{}\times\int_0^\infty d t_1 \int_0^\infty d t_2
t_1^{a_1+a_7-d/2-1} t_2^{a_2+a_8-d/2-1} (t_1+t_2)^{a_6-d/2}
\nonumber\\
&&\qquad{}K_{d/2-a_7}(t_1) K_{d/2-a_8}(t_2) K_{d/2-a_6}(t_1+t_2)\,.
\label{I2:coord}
\end{eqnarray}
If we expand the Bessel functions $K_{d/2-a_7}(t_1)$
and $K_{d/2-a_8}(t_2)$ in $t_1$ and $t_2$,
then the integrals can be calculated,
and we obtain a double series.
However, this series is not very convenient
for expansion in $\varepsilon$,
because even for convergent integrals
separate terms contain $1/\varepsilon^2$.
We have verified that the $\mathcal{O}(1)$ term
in~(\ref{prob4-03-eq2}) is reproduced
by numerical integration in~(\ref{I2:coord})
with 12 digits accuracy,
thus providing a good check.

\subsection{Integral $I_3$}
\label{S:I3}

To evaluate $I_3$
we derived a general MB representation for
$F_1(a_1,a_2,a_3,a_4,0,0,a_7,a_8,0)$, where
$I_3=F_1(1,1,1,1,0,0,1,1,0)$. To do this, we used an alpha
representation and then applied (\ref{MB}) twice in an appropriate
way. We obtained
\bea
F_1(a_1,a_2,a_3,a_4,0,0,a_7,a_8,0)
%F(\lm_1,\ldots,\lm_5)
=\frac{\Gm(a_{12}/2 + a_{347} + 2 \ep-4)
  \Gm(a_{12}/2 + a_{348} + 2 \ep-4)
  }{
4 \sqrt{\pi} \Gm((4 - a_{12} - 2 \ep)/2)
  \Gm(a_{12} + 2 a_{3478} + 4 \ep-8)}
\nn \\ && \hspace*{-137mm}
\times
\frac{\Gm(a_{12}/2 + a_{3478} + 3 \ep-6)}{ \prod \Gm(a_i)}
%\nn \\ && \hspace*{-80mm}
\frac{1}{(2\pi i)^2} \int_{-i\infty}^{+i\infty} \int_{-i\infty}^{+i\infty}
\dd z_1\dd z_2\;
\frac{\Gm(a_1/2 + z_2) \Gm(1/2 + a_1/2 + z_2)}{\Gm(1/2 - z_1)}
\nn \\ &&  \hspace*{-137mm}\times
 \Gm(a_{12}/2 + z_1 + z_2) \Gm(a_{34} + \ep -2- z_1)
 \Gm(2 - a_{12}/2 - a_4 - \ep - z_2)
\nn \\ && \hspace*{-137mm}
\Gm(-a_1/2 - z_1 - z_2) \Gm(1/2 - a_1/2 - z_1 - z_2)
\Gm(2 - a_3 - \ep + z_1 + z_2)\Gm(-z_2)
%\nn \\ &&  \hspace*{-80mm}\times
%
\;.
\label{I3MB}
\eea

As in the previous case, we preferred to calculate, instead of
$I_3$, another integral of this family, $F_1(2,2,1,1,0,0,1,1,0)$
and then obtain $I_3$ using our reduction procedure.
We have, after some changes of variables,
\bea
F_1(2,2,1,1,0,0,1,1,0)
%F(\lm_1,\ldots,\lm_5)
=\frac{\Gm(3 \ep) \Gm(1 + 2 \ep)^2}{
4 \sqrt{\pi}  \Gm(-\ep) \Gm(2 + 4 \ep) }
\nn \\ && \hspace*{-80mm}
\times
\frac{1}{(2\pi i)^2} \int_{-i\infty}^{+i\infty} \int_{-i\infty}^{+i\infty}
\dd z_1\dd z_2\;
\frac{  \Gm(3/2 + z_1)
  \Gm(1 + z_2) \Gm(3/2 + z_2)\Gm(1 + z_1)
 }{
 \Gm(5/2 + z_1 + z_2)  }
\nn \\ &&  \hspace*{-80mm}\times
\Gm(2 + \ep + z_1 + z_2)  \Gm(-1 - \ep - z_1) \Gm(-z_1) \Gm(-1 - \ep - z_2) \Gm(-z_2)
\;.
\label{I3MBa}
\eea
After the replacement
\[
\Gm(2 + \ep + z_1 + z_2) \to (1 + \ep + z_1 + z_2)\Gm(1 + \ep + z_1 + z_2)
\]
the integral can be decomposed into two integrals where the
integration can be performed using the first Barnes lemma.
Resulting onefold integrals can be also evaluated with the help of
the first and the second Barnes lemmas so that we obtain a result
in terms of gamma functions for general $d$:
\bea
I_3
=\frac{
\Gm(1/2 - \ep) \Gm(-\ep) \Gm(2 \ep)^2 \Gm(1 + \ep) \Gm(3 \ep-1)}
{4 \Gm(3/2 - \ep) \Gm(4 \ep)}
\nn \\ && \hspace*{-60mm}
\times
  \left[\psi(1/2 - \ep) +
  \psi(1 - \ep) + 2\ln 2 +2\gm_{\rm E}\right]
 \;.
\label{I3res}
\eea

\subsection{Integral $I_4$}
\label{S:I4}

To obtain a MB representation for the integrals
$F_1(a_1,a_2,0,a_4,a_5,a_6,a_7,0,0)$, with
$I_4=F_1(1,1,0,1,1,1,1,0,0)$, we replace the internal integral
over $k_2$ by a onefold MB integral, using Feynman parameters
straightforwardly. After this the resulting integral over $k_1$
and $k_3$ can be evaluated in terms of gamma functions and we
obtain a general onefold MB representation which gives
\bea
I_4
=\frac{\Gm(2 \ep)^2\Gm(3 \ep-1)}{4 \Gm(4 \ep)}
\frac{1}{2\pi i} \int_{-i\infty}^{+i\infty}
\dd z
\frac{\Gm(1 + z) \Gm(1/2 + \ep + z) \Gm(1 + \ep + z)}
{\Gm(3/2 + \ep + z) \Gm(1 - 2 \ep - z)}
\nn \\ && \hspace*{-60mm}
\times
%\frac{}
\Gm(-2 \ep - z) \Gm(-\ep - z) \Gm(-z)
%{ }
\;.
\label{I4MB}
\eea
This integral can straightforwardly be evaluated by expanding the
integrand in $\ep$, closing the integration contour to
the right and summing up resulting series. We obtain
\begin{eqnarray}
I_4 &=& - \Gamma^3(1+\varepsilon) \Biggl[
\frac{\pi^2}{9\varepsilon^2}
- \frac{6\zeta_3-5\pi^2}{9\varepsilon}
+ \frac{11}{270} \pi^4 - \frac{10}{3} \zeta_3 + \frac{19}{9} \pi^2
\nonumber\\
&&{} + \left( - \frac{8}{3} \zeta_5 + \frac{8}{9} \pi^2 \zeta_3
+ \frac{11}{54} \pi^4 - \frac{38}{3} \zeta_3 + \frac{65}{9} \pi^2
\right) \varepsilon + \mathcal{O}(\varepsilon^2) \Biggr]\,.
\label{I4res}
\end{eqnarray}

\subsection{Integral $J_3$}
\label{S:J3}

To obtain a MB representation for the integrals
$F_2(a_1,a_2,a_3,a_4,a_5,a_6,0,0,0)$, with
$J_3=F_2(1,0,1,1,1,1,0,0,0)$,
we use the following onefold MB representation
for a self-energy one-loop integral with two equal masses
which can straightforwardly be derived using Feynman parameters:
\bea
\frac{1}{i\pi^{d/2}}
\int \frac{\dd^d k }{(-k^2+m^2)^{a_1}
[-(q-k)^2+m^2]^{a_2}}
%&& \nn \\ && \hspace*{-80mm}
=\frac{1}
{\Gm(a_1)\Gm(a_2)(m^2)^{a_1+a_2+\ep-2}}
&&\nn \\  && \hspace*{-110mm}
\times
\frac{1}{2\pi i}
\int_{-i\infty}^{+i \infty} \dd z
\left(\frac{-q^2}{m^2}\right)^z
\Gm(-z)
%\nn \\  && \hspace*{-80mm}
%\times
\frac{\Gm(a_1+z)\Gm(a_2+z)\Gm(a_1+a_2+\ep-2+z)}
{\Gm(a_1+a_2+2 z)}
\; .
\label{1lmm0MB}
\eea

Writing down the subintegral over $k_2$
using (\ref{1lmm0MB}) we obtain an integral which can explicitly be
evaluated in terms of gamma functions for general $\ep$ so that we
obtain only onefold MB representation for the integral under
consideration:
%by
%(\ref{1lmm0MB}) and then apply (\ref{TI35c}) to obtain the
%following onefold MB representation:
\bea
F_2(a_1,a_2,a_3,a_4,a_5,a_6,0,0,0) &=&
\frac{  \Gm(a_1/2)}{
2\prod_{l\neq 2}\Gm(a_l) \Gm(2-a_{1}/2-\ep)
(m^2)^{a_1/2+a_{23456}-6+3\ep} }
\nn \\ &&  \hspace*{-25mm}\times
\frac{1}{2\pi i}
\int_{-i\infty}^{+i\infty}
\dd z\, \frac{\Gm(a_4 + z)\Gm(a_6 + z)
\Gm(-z)\Gm(a_1/2 + a_{235} + 2\ep-4 - z)}
{\Gm(a_{46} + 2z)\Gm(a_{12235} + 2 \ep -4- 2 z)}
\nn \\ &&  \hspace*{-25mm}\times
\Gm(2 - a_1/2 - a_2 - \ep + z)
\Gm(a_{46} + \ep-2 + z)
\nn \\ &&  \hspace*{-25mm}\times
\Gm(a_1/2 + a_{23} + \ep-2 - z)
\Gm(a_1/2 + a_{25} + \ep-2 - z)
\;.
\eea

This gives
\bea
J_3
=\frac{\pi^{3/2} }{4^\ep  \Gm(3/2 - \ep)}
\frac{1}{2\pi i} \int_{-i\infty}^{+i\infty}
\dd z
\frac{ \Gm(1 + z) \Gm(3/2 - \ep + z) \Gm(\ep + z)
}{\Gm(3/2 + z)\Gm(\ep - z) }
\nn \\ && \hspace*{-60mm}
\times
%\frac{}
\Gm(-1/2 + \ep - z) \Gm(-3/2 + 2 \ep - z) \Gm(-z)
%{ }
\;.
\label{J3MB}
\eea
For our purposes, it was sufficient to
evaluate only the leading term of this integral
(using the strategy of \cite{MB1}) in expansion in $\ep$:
\be
J_3 =%e^{-3\gm_{\rm E} \ep}
%\frac{\left(i\pi^{d/2}e^{-\gm_{\rm E} \ep}\right)^3 }
%{(v^2)^2 (m^2)^{1+3\ep}}
%\left(
-\frac{32 \pi^2}{3} +\mathcal{O}(\ep)
%\right)
\;.
\label{J3res}
\ee

\section{Decoupling for the heavy-quark field}
\label{S:DQ}

\subsection{General formulae}
\label{S:field}

It is convenient to find the relation between the bare fields
in the ``full '' and the low-energy HQET,
\begin{equation}
\tilde{Q}_0 = \left(\tilde{\zeta}_Q^0\right)^{1/2} \tilde{Q}'_0\,,
\label{field:bare}
\end{equation}
first; then
\begin{equation}
\tilde{\zeta}_Q(\mu) =
\frac{\tilde{Z}'_Q(\alpha_s'(\mu),a'(\mu))}{\tilde{Z}_Q(\alpha_s(\mu),a(\mu))}
\tilde{\zeta}_Q^0\,.
\label{field:ren}
\end{equation}
The renormalization constant $\tilde{Z}_Q$ can be reconstructed
from the three-loop anomalous dimension of the HQET heavy-quark field $\tilde{Q}$
which has been found in~\cite{MR:00} by an on-shell massive QCD calculation
and confirmed by an independent HQET calculation in~\cite{CG:03}.

The bare heavy-quark propagators in the two theories
near the mass shell ($\omega\to0$) behave as
\begin{equation}
\tilde{S}(\omega) =
\frac{\tilde{Z}_Q^{\mbox{\scriptsize os}}}{\omega} + \cdots\,,\qquad
\tilde{S}'(\omega) =
\frac{\tilde{Z}_Q^{\prime\mbox{\scriptsize os}}}{\omega} + \cdots\,,
\label{field:os}
\end{equation}
where $\tilde{Z}_Q^{\mbox{\scriptsize os}}$
and $\tilde{Z}_Q^{\prime\mbox{\scriptsize os}}$
are the on-shell renormalization constants
of the heavy-quark field in these theories.
Therefore,
\begin{equation}
\tilde{\zeta}_Q^0 =
\frac{\tilde{Z}_Q^{\mbox{\scriptsize os}}}%
{\tilde{Z}_Q^{\prime\mbox{\scriptsize os}}}\,.
\label{field:zeta0}
\end{equation}

The bare heavy-quark propagator can be written as
\begin{equation}
\tilde{S}(\omega) = \frac{1}{\omega-\tilde{\Sigma}(\omega)}\,,
\label{field:prop}
\end{equation}
where $-i\tilde{\Sigma}(\omega)$ is the sum of all
one-particle-irreducible self-energy diagrams.
Therefore,
\begin{equation}
\tilde{Z}_Q^{\mbox{\scriptsize os}} =
\frac{1}{1-\left(d\tilde{\Sigma}(\omega)/d\omega\right)_{\omega=0}}\,,
\label{field:Z}
\end{equation}
and similarly
\begin{equation}
\tilde{Z}_Q^{\prime\mbox{\scriptsize os}} =
\frac{1}{1-\left(d\tilde{\Sigma}'(\omega)/d\omega\right)_{\omega=0}}\,.
\label{field:Zprime}
\end{equation}
All diagrams for
$(d\tilde{\Sigma}'(\omega)/d\omega)_{\omega=0}$
contain no scale and hence vanish, and
\begin{equation}
\tilde{Z}_Q^{\prime\mbox{\scriptsize os}} = 1\,.
\label{field:Z1}
\end{equation}

\subsection{Bare calculation in full HQET}
\label{S:self}

Only diagrams with (at least one) $c$-quark loop contribute
to $(d\tilde{\Sigma}(\omega)/d\omega)_{\omega=0}$.
Such diagrams first appear at two loops (Fig.~\ref{F:Sigma}a).
This diagram has been calculated in~\cite{BG:95}:
\begin{equation}
\left(\frac{d\tilde{\Sigma}(\omega)}{d\omega}\right)^{(2)}_{\omega=0} =
- C_F T_F \frac{g_0^4 m_{c0}^{-4\varepsilon}}{(4\pi)^d} I_0^2
\frac{(d-1)(d-2)(d-6)}{2(d-5)(d-7)}
\label{self:res2}
\end{equation}
(of course, $g_0$ and the bare $c$-quark mass $m_{c0}$
are those of the full theory).

\FIGURE{
\begin{picture}(150,137)
\put(21,125.5){\makebox(0,0){\includegraphics{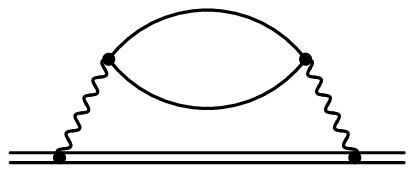}}}
\put(75,126.75){\makebox(0,0){\includegraphics{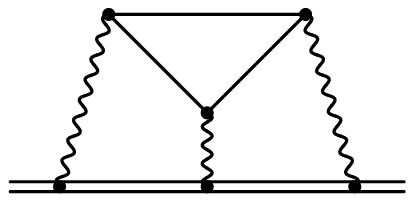}}}
\put(129,125.5){\makebox(0,0){\includegraphics{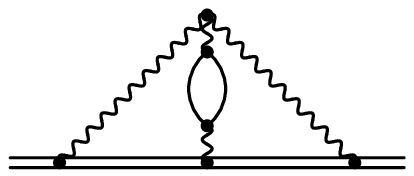}}}
\put(21,98.5){\makebox(0,0){\includegraphics{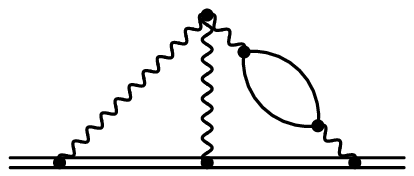}}}
\put(75,98.5){\makebox(0,0){\includegraphics{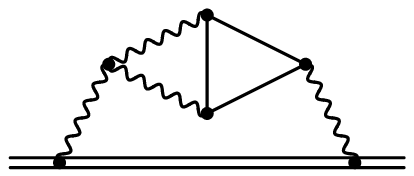}}}
\put(129,96.5){\makebox(0,0){\includegraphics{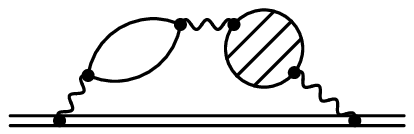}}}
\put(26,69.375){\makebox(0,0){\includegraphics{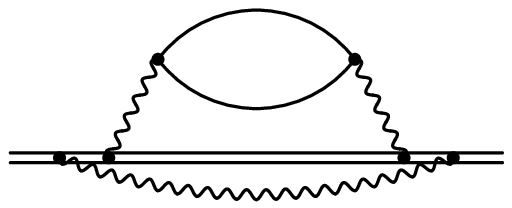}}}
\put(75,69.375){\makebox(0,0){\includegraphics{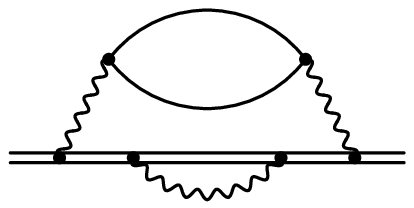}}}
\put(124,69.375){\makebox(0,0){\includegraphics{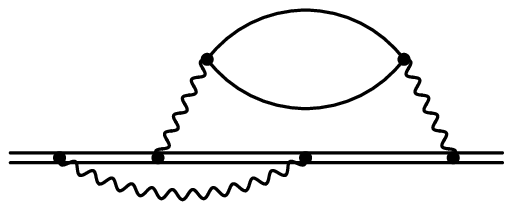}}}
\put(21,38){\makebox(0,0){\includegraphics{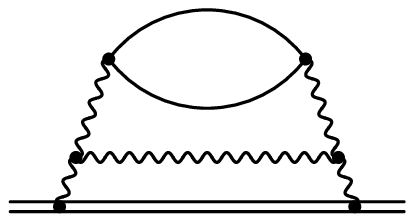}}}
\put(75,35.5){\makebox(0,0){\includegraphics{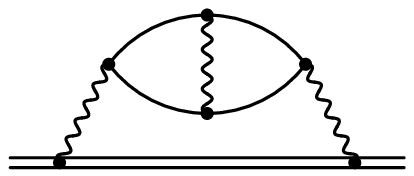}}}
\put(129,35.5){\makebox(0,0){\includegraphics{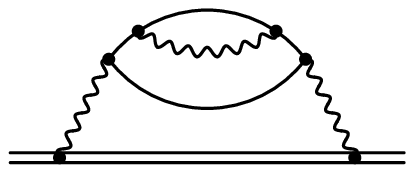}}}
\put(21,11){\makebox(0,0){\includegraphics{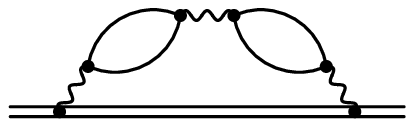}}}
\put(21,112){\makebox(0,0)[b]{a}}
\put(75,112){\makebox(0,0)[b]{b}}
\put(129,112){\makebox(0,0)[b]{c}}
\put(21,85){\makebox(0,0)[b]{d}}
\put(75,85){\makebox(0,0)[b]{e}}
\put(129,85){\makebox(0,0)[b]{f}}
\put(26,54){\makebox(0,0)[b]{g}}
\put(75,54){\makebox(0,0)[b]{h}}
\put(124,54){\makebox(0,0)[b]{i}}
\put(21,22){\makebox(0,0)[b]{j}}
\put(75,22){\makebox(0,0)[b]{k}}
\put(129,22){\makebox(0,0)[b]{l}}
\put(21,0){\makebox(0,0)[b]{m}}
\put(95,114){\makebox(0,0)[br]{+s}}
\put(41,87){\makebox(0,0)[br]{+s}}
\put(95,87){\makebox(0,0)[br]{+s}}
\put(149,87){\makebox(0,0)[br]{+s}}
\put(149,56){\makebox(0,0)[br]{+s}}
\put(149,24){\makebox(0,0)[br]{+s}}
\end{picture}
\caption{Diagrams for
$(d\tilde{\Sigma}(\omega)/d\omega)_{\omega=0}$}
\label{F:Sigma}}

Three-loop diagrams contributing
to $(d\tilde{\Sigma}(\omega)/d\omega)_{\omega=0}$
are shown in Fig.~\ref{F:Sigma}b--m,
where the shaded blob means the sum
of the massless one-loop insertions (light quark, gluon, ghost),
and ``+s'' means that there is also a mirror-symmetric diagram.
The diagram Fig.~\ref{F:Sigma}i has three HQET denominators
depending on only two loop momenta;
these denominators are linearly dependent,
and we can eliminate one HQET line by taking partial fractions.
After that, all diagrams of Fig.~\ref{F:Sigma}b--j
can be expressed via scalar integrals of class 1 (Sect.~\ref{S:1});
those of Fig.~\ref{F:Sigma}k--m can be expressed via scalar integrals
of class 2 (Sect.~\ref{S:2}).

The three-loop result can be written in the form
\begin{equation}
\left(\frac{d\tilde{\Sigma}(\omega)}{d\omega}\right)^{(3)}_{\omega=0} =
C_F T_F \frac{g_0^6 m_{c0}^{-6\varepsilon}}{(4\pi)^{3d/2}}
\sum_{i=1}^4 \sum_{j=1}^5 C_i \bar{I}_j s_{ij}\,,
\label{self:res3}
\end{equation}
with the colour structures
\begin{equation}
C_1 = C_F\,,\qquad
C_2 = C_A\,,\qquad
C_3 = T_F n_l\,,\qquad
C_4 = T_F\,,
\label{self:colour}
\end{equation}
and the integral structures
\begin{eqnarray}
\bar{I}_1 &=&
\frac{(d-2)^2}{24(d-3)(d-4)^2(d-5)^3(d-6)(d-7)(d-8)(d-10)}
I_0^3\,,
\nonumber\\
\bar{I}_2 &=&
\frac{(3d-8)(3d-10)}{16(d-3)(d-4)(d-5)^2(2d-7)(2d-9)(2d-11)}
I_1\,,
\nonumber\\
\bar{I}_3 &=& \frac{3d-10}{(d-5)(2d-9)(2d-11)} I_3\,,\qquad
\bar{I}_4 = \frac{3d-10}{(d-5)(2d-9)(2d-11)} I_4\,,
\nonumber\\
\bar{I}_5 &=& \frac{(3d-8)(3d-10)}{16(d-4)^2(d-6)(d-8)(d-10)} I_5\,.
\label{self:int}
\end{eqnarray}
All the non-zero coefficients are:
\begin{eqnarray}
s_{11} &=& 6 (d-5)^2 (d-10)
(2 d^8 - 75 d^7 + 1212 d^6 - 11042 d^5 + 62070 d^4 - 220131 d^3
\nonumber\\
&&{} + 478504 d^2 - 576108 d + 287408)\,,
\nonumber\\
s_{12} &=& 2 (d-2) (d-3) (d-5)^2 (2d-11)
(2 d^4 - 29 d^3 + 157 d^2 - 366 d + 288)\,,
\nonumber\\
s_{15} &=& - 2 s_{25} = - 2 (d-10) (d^4 - 20 d^3 + 141 d^2 - 398 d + 328)\,,
\nonumber\\
s_{21} &=& (d-7) (d-10)
(15 d^8 - 555 d^7 + 8858 d^6 - 79542 d^5 + 438503 d^4 - 1514103 d^3
\nonumber\\
&&{} + 3179372 d^2 - 3678156 d + 1767408)\,,
\nonumber\\
s_{22} &=& - (d-3)
(2 d^8 - 63 d^7 + 864 d^6 - 6725 d^5 + 32377 d^4 - 98156 d^3 + 181403 d^2
\nonumber\\
&&{} - 184490 d + 77536)
+ 2 (d-1) (d-3)^2 (d-4) (d-5)^3 (1-a_0)\,,
\nonumber\\
s_{23} &=& - s_{24} = 2 (2 d^2 - 19 d + 47)\,,\qquad
s_{32} = 16 (d-2) (d-3) (d-4) (d-5)^3\,,
\nonumber\\
s_{41} &=& 12 (d-4) (d-5)^2
(9 d^6 - 277 d^5 + 3387 d^4 - 20943 d^3 + 68428 d^2 - 110236 d + 66352)\,,
\nonumber\\
s_{45} &=& 4 (d-4) (d^3 - 15 d^2 + 54 d - 8)\,.
\label{self:coef}
\end{eqnarray}

The coefficient of $C_A I_1$ is not gauge invariant.
Therefore, the on-shell renormalization constant
of the HQET heavy-quark field~(\ref{field:Z})
is not gauge-invariant, starting from three loops.
The same phenomenon has been observed in QCD~\cite{MR:00}.
In the abelian case, the on-shell renormalization constant
is gauge invariant to all orders of perturbation theory;
the gauge dependence only appears in the non-abelian
colour structure $C_A$.

\subsection{Renormalized decoupling coefficient}
\label{S:DQren}

Now we calculate the renormalized decoupling coefficient~(\ref{field:ren}).
To this end, we re-express $\tilde{Z}_Q$ via $g_0^2$, $a_0$;
$\tilde{Z}'_Q$ via $g_0^{\prime2}$, $a'_0$,
which are re-expressed via $g_0^2$, $a_0$
using the bare decoupling relations;
combine all factors in~(\ref{field:ren});
and, finally, re-express the result via $\alpha_s(\mu)$, $a(\mu)$
and expand the coefficients in $\varepsilon$.
We arrive at
\begin{eqnarray}
&&\tilde{\zeta}_Q(\mu) = 1
+ C_F T_F \left( 2 L^2 - \frac{16}{3} L + \frac{52}{9} \right)
\left(\frac{\alpha_s(\mu)}{4\pi}\right)^2
\nonumber\\
&&{} + C_F T_F \Biggl\{
\left[ \frac{1}{3} \left( a(\mu) + \frac{44}{3} \right) C_A
- \frac{16}{9} T_F (n_l + 2) \right] L^3
\nonumber\\
&&\qquad{} + \left[ - 18 C_F
- \frac{1}{3} \left( \frac{13}{2} a(\mu) - 31 \right) C_A
+ \frac{64}{9} T_F \right] L^2
\nonumber\\
&&\qquad{} + \left[ 6 (8 \zeta_3 - 1) C_F
+ \left( \frac{121}{18} a(\mu) - 48 \zeta_3 + \frac{155}{9} \right) C_A
- \frac{16}{27} T_F (36 n_l + 31) \right] L
\nonumber\\
&&\qquad{}
+ \left( 16 B_4 - \frac{4}{5} \pi^4 + 106 \zeta_3 - \frac{475}{6} \right) C_F
\nonumber\\
&&\qquad{} + \left[
\frac{1}{3} \left( 8 \zeta_3 - \frac{2387}{72} \right) a(\mu)
- 8 B_4 + \frac{52}{45} \pi^4 - \frac{413}{9} \zeta_3 + \frac{262}{243}
\right] C_A
\nonumber\\
&&\qquad{} - \frac{32}{9} \left( 4 \zeta_3 - \frac{179}{27} \right) T_F n_l
+ \frac{8}{9} \left( 28 \zeta_3 - \frac{395}{27} \right) T_F \Biggr\}
\left(\frac{\alpha_s(\mu)}{4\pi}\right)^3 + \cdots
\label{DQren:res}
\end{eqnarray}
where
\begin{equation}
L = 2 \ln \frac{\mu}{m_c(\mu)}
\label{DQren:L}
\end{equation}
($m_c(\mu)$ is the $\overline{\mathrm{MS}}$ renormalized mass),
and $B_4$ is given by~(\ref{Simple:B4}).

\section{Decoupling for the heavy--light current}
\label{S:Dj}

\subsection{General formulae}
\label{S:current}

The bare heavy--light current
$\tilde{\jmath}_0=\bar{q}_0 \Gamma \tilde{Q}_0$
in the ``full'' HQET is related to
the corresponding operator $\tilde{\jmath}'_0$
in the low-energy theory by
\begin{equation}
\tilde{\jmath}_0 = \tilde{\zeta}_j^0 \tilde{\jmath}'_0\,,
\label{current:bare}
\end{equation}
up to $1/m_c$ corrections.
The renormalized operators are related by~(\ref{Intro:Decj}),
where
\begin{equation}
\tilde{\zeta}_j(\mu) =
\frac{\tilde{Z}'_j(\alpha_s'(\mu))}{\tilde{Z}_j(\alpha_s(\mu))}
\tilde{\zeta}_j^0\,.
\label{current:ren}
\end{equation}
The renormalization constant $\tilde{Z}_j$ can be reconstructed
from the three-loop anomalous dimension of the HQET current $\tilde{\jmath}$
which has been calculated in~\cite{CG:03}.

\FIGURE{
\begin{picture}(32,12)
\put(16,6){\makebox(0,0){\includegraphics{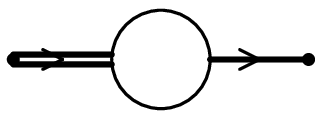}}}
\put(6,4){\makebox(0,0)[t]{$\omega$}}
\put(26,4){\makebox(0,0)[t]{$p$}}
\put(16,6){\makebox(0,0){$\tilde{\Gamma}$}}
\end{picture}
\caption{Green function}
\label{F:Green}}

The Green function of $\tilde{\jmath}_0$, $\bar{\tilde{Q}}_0$,
and $q_0$ can be written as
$S(p) \tilde{\Gamma}(p,\omega) \Gamma \tilde{S}(\omega)$
(Fig.~\ref{F:Green}),
where $\tilde{\Gamma}(p,\omega) \Gamma$ is the proper vertex ---
the sum of all one-particle-irreducible diagrams
(here $\Gamma$ is the Dirac matrix in the current;
due to the HQET Feynman rules, no $\gamma$-matrices can appear
to the right from $\Gamma$).
The Green functions in the two theories are related by
\begin{equation}
S(p) \tilde{\Gamma}(p,\omega) \Gamma \tilde{S}(\omega) =
\left( \tilde{\zeta}_Q^0 \zeta_q^0 \right)^{1/2} \tilde{\zeta}_j^0
S'(p) \tilde{\Gamma}'(p,\omega) \Gamma \tilde{S}'(\omega)
+ \mathcal{O}\left(\frac{p,\omega}{m_c}\right)\,.
\label{current:Green}
\end{equation}
Recalling the relations
\begin{equation}
S(p) = \zeta_q^0 S'(p)\,,\qquad
\tilde{S}(\omega) = \tilde{\zeta}_Q^0 \tilde{S}'(\omega)
\label{current:Green2}
\end{equation}
(up to $1/m_c$ corrections), we obtain
\begin{equation}
\tilde{\Gamma}(p,\omega) =
\left( \tilde{\zeta}_Q^0 \zeta_q^0 \right)^{-1/2} \tilde{\zeta}_j^0
\tilde{\Gamma}'(p,\omega)
+ \mathcal{O}\left(\frac{p,\omega}{m_c}\right)\,.
\label{current:Vert}
\end{equation}

\noindent
It is most convenient to use the point $p=0$, $\omega=0$,
then power corrections may be omitted.
The vertices $\tilde{\Gamma}(0,0)$, $\tilde{\Gamma}'(0,0)$
can have Dirac structures $1$ and $\rlap/v$;
if the light quark $q$ is massless,
the number of $\gamma$-matrices in them is always even,
and the second structure does not appear.
So, these vertices are scalar, and
\begin{equation}
\tilde{\zeta}_j^0 =
\left( \tilde{\zeta}_Q^0 \zeta_q^0 \right)^{1/2}
\frac{\tilde{\Gamma}(0,0)}{\tilde{\Gamma}'(0,0)}\,.
\label{current:zeta0}
\end{equation}
All loop corrections to $\tilde{\Gamma}'(0,0)$
contain no scale and hence vanish:
\begin{equation}
\tilde{\Gamma}'(0,0) = 1\,.
\label{current:G1}
\end{equation}
Therefore,
\begin{equation}
\tilde{\zeta}_j^0 =
\left( \tilde{Z}_Q^{\mbox{\scriptsize os}} Z_q^{\mbox{\scriptsize os}}
\right)^{1/2}
\tilde{\Gamma}(0,0)\,.
\label{current:z0}
\end{equation}
The right-hand side is an on-shell matrix element
of the gauge-invariant operator $\tilde{\jmath}_0$,
and hence is gauge-invariant;
the quantities $\tilde{\Gamma}(0,0)$,
$\tilde{Z}_Q^{\mbox{\scriptsize os}}$, $Z_q^{\mbox{\scriptsize os}}$
taken separately are not gauge-invariant
starting from three loops.

\subsection{Bare calculation in full HQET}
\label{S:vert}

\FIGURE{
\includegraphics{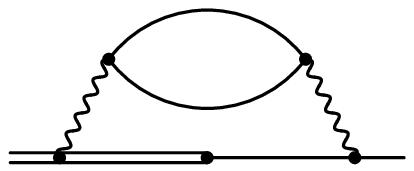}
\caption{Diagram for $\tilde{\Lambda}(0,0)$}
\label{F:Lambda}}

Only diagrams with (at least one) $c$-quark loop contribute to
$\tilde{\Lambda}(0,0)=\tilde{\Gamma}(0,0)-1$.
Such diagrams first appear at two loops (Fig.~\ref{F:Lambda}).
As we know, the result is scalar;
so, we can take $\frac{1}{4}\mathop{\mathrm{Tr}}$
of the $\gamma$-matrices associated with the light-quark line.
In this diagram, there is only one light-quark propagator $S(k)$
and the vertex, and taking the trace gives $k^\mu$.
The $c$-quark loop inserted into the gluon propagator is transverse,
and the diagram vanishes~\cite{BG:95}:
\begin{equation}
\tilde{\Lambda}^{(2)}(0,0) = 0\,.
\label{vert:res2}
\end{equation}

Three-loop diagrams for $\tilde{\Lambda}(0,0)$ can be obtained
from those in Fig.~\ref{F:Sigma}b--m by inserting
the heavy--to--light vertex into all possible places
along the heavy-quark line;
of course, mirror-symmetric diagrams
denoted by ``+s'' in Fig.~\ref{F:Sigma}
should be considered separately.
A number of diagrams with a single light-quark propagator
vanish for the same reason as the two-loop diagram
(Fig.~\ref{F:Lambda}).
For some diagrams with three HQET lines,
we use partial fractioning.
After that, all diagrams produced from those
of Fig.~\ref{F:Sigma}b--j are expressed via scalar integrals
of class 1 (Sect.~\ref{S:1});
diagrams produced from Fig.~\ref{F:Sigma}k--m
are expressed via scalar integrals of class 2 (Sect.~\ref{S:2}).

The three-loop result can be written in the form
\begin{equation}
\tilde{\Lambda}^{(3)}(0,0) =
C_F T_F \frac{g_0^6 m_{c0}^{-6\varepsilon}}{(4\pi)^{3d/2}}
\sum_{i=1}^2 \sum_{j=2}^4 C_i \bar{I}_j \lambda_{ij}
\label{vert:res3}
\end{equation}
(see~(\ref{self:colour}) and~(\ref{self:int})).
All the non-zero coefficients are:
\begin{eqnarray}
\lambda_{12} &=& - 2 (d-2)^2 (d-4)^2 (d-5)^3\,,\qquad
\lambda_{13} = \lambda_{14} = 2 (d-2) (d-5)^2\,,
\nonumber\\
\lambda_{22} &=& (d-3)^2 (d-4) (d-5)^3 (d+2)
- 2 (d-1) (d-3)^2 (d-4) (d-5)^3 (1-a_0)\,,
\nonumber\\
\lambda_{23} &=& - (d-5)^2\,,\qquad
\lambda_{24} = - (d-3) (d-5)^2\,.
\label{vert:coef}
\end{eqnarray}
Here again the coefficient of $C_A I_1$ is not gauge invariant.
The on-shell matrix element~(\ref{current:z0}) is gauge-invariant;
this is a strong check of our calculation.

\subsection{Renormalized decoupling coefficient}
\label{S:Djren}

Now we calculate the renormalized decoupling coefficient~(\ref{current:ren}).
To this end, we re-express $\alpha_s(\mu)$ in $\tilde{Z}_Q$ via $g_0^2$;
$\alpha_s'(\mu)$ in $\tilde{Z}'_Q$ via $g_0^{\prime2}$,
which is, in its turn, re-expressed via $g_0^2$
using the bare decoupling coefficient $\zeta_\alpha^0$;
combine all factors in~(\ref{current:ren});
and, finally, re-express the result via $\alpha_s(\mu)$
and expand the coefficients in $\varepsilon$.
We arrive at
\begin{eqnarray}
&&\tilde{\zeta}_j(\mu) = 1
+ C_F T_F \left( L^2 - \frac{5}{3} L + \frac{89}{36} \right)
\left(\frac{\alpha_s(\mu)}{4\pi}\right)^2
\nonumber\\
&&{} + C_F T_F \Biggl\{
\frac{2}{9} \left[ 11 C_A - 4 T_F (n_l+2) \right] L^3
+ \frac{1}{9} \left[ 16 (\pi^2-6) C_F - (4\pi^2-39) C_A + 20 T_F \right] L^2
\nonumber\\
&&\qquad{} + \frac{1}{3} \biggl[ \left(88 \zeta_3 - \frac{112}{9} \pi^2 - \frac{173}{6}\right) C_F
- \left(76 \zeta_3 - \frac{28}{9} \pi^2 - \frac{401}{6} \right) C_A
\nonumber\\
&&\qquad\qquad{} - \frac{2}{3} T_F \left(53 n_l + \frac{124}{3} \right) \biggr] L
\nonumber\\
&&\qquad{} + \left( 8 B_4 - \frac{86}{405} \pi^4 + \frac{1427}{27} \zeta_3
+ \frac{1600}{243} \pi^2 - \frac{7219}{162} \right) C_F
\nonumber\\
&&\qquad{} - \left( 4 B_4 - \frac{43}{81} \pi^4 + \frac{1471}{54} \zeta_3
+ \frac{400}{243} \pi^2 + \frac{3845}{486} \right) C_A
\nonumber\\
&&\qquad{} - \frac{2}{9} \left( 32 \zeta_3 - \frac{1327}{27} \right) T_F n_l
+ \frac{1}{9} \left( 112 \zeta_3 - \frac{1685}{27} \right) T_F \Biggr\}
\left(\frac{\alpha_s(\mu)}{4\pi}\right)^3 + \cdots
\label{Djren:zeta}
\end{eqnarray}
Here $L$ is given by~(\ref{DQren:L}) and $B_4$ by~(\ref{Simple:B4}).
This is our main result.
The two-loop part agrees with~\cite{G:98}.

At the normalization scale $\mu_c$
defined as the root of the equation
\begin{equation}
m_c(\mu_c) = \mu_c
\label{Djren:muc}
\end{equation}
we have, for the physical $SU(3)$ colour group,
\begin{eqnarray}
\tilde{\zeta}_j(\mu_c) &=& 1
+ \frac{89}{864} \left(\frac{\alpha_s(\mu_c)}{\pi}\right)^2
\nonumber\\
&&{} + \frac{1}{9} \biggl[ - 2 \Li4\left(\frac{1}{2}\right)
- \frac{1}{12} \ln^2 2 \left( \ln^2 2 - \pi^2 \right)
+ \frac{427}{3240} \pi^4 - \frac{815}{1728} \zeta_3 + \frac{175}{486} \pi^2
- \frac{877}{108}
\nonumber\\
&&\hphantom{{}+\frac{1}{9}\biggl[\biggr.}
- \frac{1}{3} \left( \zeta_3 - \frac{1327}{864} \right) n_l \biggr]
\left(\frac{\alpha_s(\mu_c)}{\pi}\right)^3
\nonumber\\
&\approx& 1 + 0.1030  \left(\frac{\alpha_s(\mu_c)}{\pi}\right)^2
+ (0.7828 + 0.0124 n_l) \left(\frac{\alpha_s(\mu_c)}{\pi}\right)^3\,,
\label{Djren:num}
\end{eqnarray}
where the number of light flavours $n_l=3$ includes neither $b$ nor $c$.

It is easy to re-write~(\ref{Djren:zeta})
via the on-shell mass $m_c^{\mbox{\scriptsize os}}$
instead of the $\overline{\mathrm{MS}}$ mass $m_c(\mu)$.
Substituting the well-known one-loop relation between them,
we obtain
\begin{eqnarray}
&&\tilde{\zeta}_j(\mu) = 1
+ C_F T_F \left( L_{\mbox{\scriptsize os}}^2 - \frac{5}{3} L_{\mbox{\scriptsize os}}
+ \frac{89}{36} \right)
\left(\frac{\alpha_s(\mu)}{4\pi}\right)^2
\nonumber\\
&&{} + C_F T_F \Biggl\{
\frac{2}{9} \left[ 11 C_A - 4 T_F (n_l+2) \right] L_{\mbox{\scriptsize os}}^3
+ \frac{1}{9} \left[ 4 (4\pi^2+3) C_F - (4\pi^2-39) C_A + 20 T_F \right]
L_{\mbox{\scriptsize os}}^2
\nonumber\\
&&\qquad{} + \frac{1}{3} \biggl[ \left(88 \zeta_3 - \frac{112}{9} \pi^2
- \frac{65}{6}\right) C_F
- \left(76 \zeta_3 - \frac{28}{9} \pi^2 - \frac{401}{6} \right) C_A
\nonumber\\
&&\qquad\qquad{} - \frac{2}{3} T_F \left(53 n_l + \frac{124}{3} \right) \biggr]
L_{\mbox{\scriptsize os}}
\nonumber\\
&&\qquad{} + \left( 8 B_4 - \frac{86}{405} \pi^4 + \frac{1427}{27} \zeta_3
+ \frac{1600}{243} \pi^2 - \frac{9379}{162} \right) C_F
\nonumber\\
&&\qquad{} - \left( 4 B_4 - \frac{43}{81} \pi^4 + \frac{1471}{54} \zeta_3
+ \frac{400}{243} \pi^2 + \frac{3845}{486} \right) C_A
\nonumber\\
&&\qquad{} - \frac{2}{9} \left( 32 \zeta_3 - \frac{1327}{27} \right) T_F n_l
+ \frac{1}{9} \left( 112 \zeta_3 - \frac{1685}{27} \right) T_F \Biggr\}
\left(\frac{\alpha_s(\mu)}{4\pi}\right)^3 + \cdots
\label{Djren:os}
\end{eqnarray}
where
\begin{equation}
L_{\mbox{\scriptsize os}} = 2 \ln \frac{\mu}{m_c^{\mbox{\scriptsize os}}}\,.
\label{Djren:Los}
\end{equation}
At $\mu=m_c^{\mbox{\scriptsize os}}$ we have,
for the physical $SU(3)$ colour group,
\begin{eqnarray}
\tilde{\zeta}_j(m_c^{\mbox{\scriptsize os}}) &=& 1
+ \frac{89}{864} \left(\frac{\alpha_s(m_c^{\mbox{\scriptsize os}})}{\pi}\right)^2
\nonumber\\
&&{} + \frac{1}{9} \biggl[ - 2 \Li4\left(\frac{1}{2}\right)
- \frac{1}{12} \ln^2 2 \left( \ln^2 2 - \pi^2 \right)
+ \frac{427}{3240} \pi^4 - \frac{815}{1728} \zeta_3 + \frac{175}{486} \pi^2
- \frac{1057}{108}
\nonumber\\
&&\hphantom{{}+\frac{1}{9}\biggl[\biggr.}
- \frac{1}{3} \left( \zeta_3 - \frac{1327}{864} \right) n_l \biggr]
\left(\frac{\alpha_s(m_c^{\mbox{\scriptsize os}})}{\pi}\right)^3
\nonumber\\
&\approx& 1 + 0.1030  \left(\frac{\alpha_s(m_c^{\mbox{\scriptsize os}})}{\pi}\right)^2
+ (0.5976 + 0.0124 n_l)
\left(\frac{\alpha_s(m_c^{\mbox{\scriptsize os}})}{\pi}\right)^3\,.
\label{Djren:numos}
\end{eqnarray}

\section{Conclusion}
\label{S:Conc}

The results on decoupling $c$-quark loops in the $b$-quark HQET
obtained in the present paper can be used to improve the accuracy
of extracting $f_B$ from lattice HQET simulations.
It requires the following steps (at least conceptually;
in practice, some of them can be grouped together):
\begin{itemize}
\item Matching the lattice HQET to the continuum HQET
in the $\overline{\mbox{MS}}$ scheme at a low scale $\mu\sim1/a$,
where $a$ is the lattice spacing. It can be done using the lattice
perturbation theory, or non-perturbatively.
\item Running up to $\mu\sim m_c$ in the low-energy HQET
with 3 active flavours.
\item Matching the low-energy HQET to HQET with 4 active flavours.
This is the subject of the present paper.
\item Running up to $\mu\sim m_b$ in HQET with 4 active flavours.
\item Matching to QCD with 5 active flavours, including $b$.
\end{itemize}
Currently, all steps
(except the first one, which is lattice-specific)
can be done at the next-to-next-to-leading order:
the QCD/HQET matching is known at two loops~\cite{BG:95,G:98},
and the HQET heavy--light current anomalous dimension
(which determines running in both HQETs) ---
at three loops~\cite{CG:03}.
The present paper is a first step
towards the N$^3$L order calculation:
the step 3 can now be done with the three-loop accuracy.
To complete this program, we need the QCD/HQET matching coefficients
at three loops and the anomalous dimension at four loops.
The first task does not seem impossible:
the QCD/HQET matching at three loops can be calculated
using the methods of~\cite{MR:00}.
Prospects of obtaining the four-loop anomalous dimension
are doubtful.
However, the intervals of $\mu$
(from $1/a$ to $m_c$ and from $m_c$ to $m_b$)
are not really wide.
Though formally the N$^3$L order calculation requires
four-loop running, its effect is likely to be small,
and three-loop running should be sufficient.

We have also obtained the generic formula~(\ref{Djll:res})
for the three-loop decoupling coefficient
for the flavour-nonsinglet QCD current
with $n$ antisymmetrized $\gamma$-matrices.
The expressions for $n=0$, 1, 3, 4 can be obtained from published results;
the formula for $n=2$ is new.

The results of the present paper in a computer-readable form
can be found at~\cite{progdata}.

The method of calculation of three-loop on-shell HQET diagrams
with massive-quark loops can be applied to other physical problems.
We hope to consider them in a future publication.

\acknowledgments

We are grateful to K.G.~Chetyrkin for his great help.
The work of A.G.\ was supported by the DFG through SFB/TR 9.
The work of A.S.\ was supported by the Russian Foundation for Basic
Research through  grant 05-01-00988.
%and by CRDF through grant RM1-2543.
The work of V.S.\ was supported by the Russian Foundation for Basic
Research through grant 05-02-17645.

\appendix
\section{Decoupling in QCD}
\label{S:QCD}

\subsection{Decoupling for the light-quark field}
\label{S:Dq}

In order to find the decoupling coefficient for the heavy--light current,
we need the decoupling coefficients for both the heavy-quark field
(Sect.~\ref{S:DQ}) and the light-quark one.
The last coefficient has been calculated in~\cite{CKS:98};
however, only the result for $N_c=3$ has been presented in the paper.
Here we re-calculate this quantity.
At $N_c=3$, our result coincides with~\cite{CKS:98}%
\footnote{K.G.~Chetyrkin has kindly provided an unpublished three-loop formula
for $\zeta_q(\mu)$ for the $SU(N_c)$ colour group;
our result coincides with it.}.

Similarly to Sect.~\ref{S:DQ}, the bare decoupling coefficient
for the light-quark field is given by
\begin{equation}
\zeta_q^0 = \frac{Z_q^{\mbox{\scriptsize os}}}{Z_q^{\prime\mbox{\scriptsize os}}}\,,
\label{Dq:zeta0}
\end{equation}
where
\begin{equation}
Z_q^{\mbox{\scriptsize os}} = \frac{1}{1-\Sigma_V(0)}\,,\qquad
{Z_q^{\prime\mbox{\scriptsize os}}} = \frac{1}{1-\Sigma'_V(0)} = 1\,,
\label{Dq:Zq}
\end{equation}
and the massless-quark self-energy is $\Sigma(p)=\rlap/p\Sigma_V(p^2)$.
The renormalized decoupling coefficient is
\begin{equation}
\zeta_q(\mu) =
\frac{Z'_q(\alpha_s'(\mu),a'(\mu))}{Z_q(\alpha_s(\mu),a(\mu))}
\zeta_q^0\,.
\label{Dq:zeta}
\end{equation}

Only diagrams with a massive quark loop contribute to $\Sigma_V(0)$;
they can be obtained from Fig.~\ref{F:Sigma} by replacing the HQET line
by a massless quark line.
We made minimal replacements in the code calculating the HQET self-energy,
so that the correctness of the light-quark result provides
an additional check of the HQET calculation.
The result is
\begin{eqnarray}
\Sigma_V(0) &=& C_F T_F \frac{g_0^4 m_{c0}^{-4\varepsilon}}{(4\pi)^d} I_0^2
\frac{(d-1)(d-2)(d-4)(d-6)}{2d(d-5)(d-7)}
\nonumber\\
&&{} - C_F T_F \frac{g_0^6 m_{c0}^{-6\varepsilon}}{(4\pi)^{3d/2}}
\frac{d-4}{d} \sum_{i=1}^4 \sum_{j=1,2,5} C_i \bar{I}_j \sigma_{ij}
+ \cdots\,,
\label{Dq:Sigma}
\end{eqnarray}
where all non-zero coefficients are
\begin{eqnarray}
\sigma_{11} &=& s_{11}\,,\qquad
\sigma_{15} = - 2 \sigma_{25} = s_{15}\,,\qquad
\sigma_{32} = s_{32}\,,\qquad
\sigma_{41} = s_{41}\,,\qquad
\sigma_{45} = s_{45}\,,
\nonumber\\
\sigma_{12} &=& 2 (d-2) (d-3) (d-5)^2
(4 d^5 - 83 d^4 + 673 d^3 - 2646 d^2 + 4952 d - 3368)\,,
\nonumber\\
\sigma_{21} &=& 3 (d-5) (d-7) (d-10)
(5 d^7 - 156 d^6 + 2040 d^5 - 14470 d^4 + 59897 d^3 - 143640 d^2
\nonumber\\
&&{} + 182340 d - 92176)\,,
\nonumber\\
\sigma_{22} &=& - (d-3) (d-5)
(2 d^7 - 53 d^6 + 588 d^5 - 3557 d^4 + 12727 d^3 - 26983 d^2 + 31224 d
\nonumber\\
&&{} - 14932)
- 2 d (d-1) (d-3)^2 (d-5)^3 (1-a_0)
\label{Dq:coef}
\end{eqnarray}
(see~(\ref{self:coef})).
The on-shell renormalization constant of the light-quark field~(\ref{Dq:Zq})
is not gauge-invariant, starting from three loops;
the $a_0$-dependent term in it is the same as
in the on-shell renormalization constant of the HQET field~(\ref{field:Z}),
see~(\ref{self:res3}) and~(\ref{self:coef}).

Now we calculate the renormalized decoupling coefficient~(\ref{Dq:zeta}):
\begin{eqnarray}
&&\zeta_q(\mu) = 1
+ C_F T_F \left( 2 L - \frac{5}{6} \right)
\left(\frac{\alpha_s(\mu)}{4\pi}\right)^2
\nonumber\\
&&{} + C_F T_F \Biggl\{ \frac{1}{3} C_A a(\mu) L^3
+ \left[ 2 C_F - \left( \frac{13}{6} a(\mu) + 1 \right) C_A
- \frac{8}{3} T_F \right] L^2
\nonumber\\
&&\qquad{} + \left[ - 15 C_F
+ \frac{1}{9} \left( \frac{121}{2} a(\mu) + 232 \right) C_A
- \frac{20}{9} T_F n_l \right] L
\nonumber\\
&&\qquad{} + \left( 8 \zeta_3 + \frac{155}{18} \right) C_F
+ \left[ \frac{1}{3} \left( 8 \zeta_3 - \frac{2387}{72} \right) a(\mu)
- 8 \zeta_3 - \frac{1187}{81} \right] C_A
\nonumber\\
&&\qquad{} - \frac{70}{81} T_F (2 n_l + 1) \Biggr\}
\left(\frac{\alpha_s(\mu)}{4\pi}\right)^3 + \cdots
\label{Dq:res}
\end{eqnarray}

\subsection{Decoupling for light--light currents}
\label{S:Djll}

Here we shall consider decoupling for the flavour non-singlet QCD currents
with $n$ antisymmetrized $\gamma$-matrices
\begin{equation}
j_0 = \bar{q}_0 \Gamma \tau q_0\,,\qquad
\Gamma = \gamma^{[\alpha_1} \cdots \gamma^{\alpha_n]}\,,
\label{Djll:j}
\end{equation}
where $\tau$ is a flavour matrix with $\mathop{\mathrm{Tr}}\tau=0$,
for an arbitrary $n$ at three loops
(the two-loop result has been obtained in~\cite{G:98}).

The proper vertex $\Gamma(p,p')=\Gamma+\Lambda(p,p')$ at $p=p'=0$
has the structure $\Gamma(0,0)=\Gamma_n\Gamma$,
where $\Gamma$ is the Dirac matrix in~(\ref{Djll:j}),
and $\Gamma_n=1+\Lambda_n$ is scalar.
Similarly to Sect.~\ref{S:Dj}, the bare decoupling coefficient is
\begin{equation}
\zeta_n^0 = Z_q^{\mbox{\scriptsize os}} \Gamma_n\,,
\label{Djll:zeta0}
\end{equation}
because $Z_q^{\prime\mbox{\scriptsize os}}=1$, $\Gamma_n'=1$.
This is nothing but the on-shell matrix element of the current in the full theory,
and it must be gauge invariant.
The renormalized decoupling coefficient is
\begin{equation}
\zeta_n(\mu) =
\frac{Z'_n(\alpha_s'(\mu))}{Z_n(\alpha_s(\mu))}
\zeta_n^0\,,
\label{Djll:zeta}
\end{equation}
where $j_0=Z_n(\alpha_s(\mu))j(\mu)$,
and $Z_n$ can be reconstructed from the anomalous dimension $\gamma_n$,
which has been calculated at three loops for a generic $n$ in~\cite{Gr:00}.

The diagrams for $\Lambda^{(3)}(0,0)$ can be obtained from the HQET vertex diagrams
(Sect.~\ref{S:vert}) by replacing the HQET line by a massless quark line.
To find the result for an arbitrary $n$,
we follow the method used in~\cite{BG:95}.
If we make no assumptions about the properties of the matrix $\Gamma$,
then the three-loop vertex has the structure
\begin{equation}
\Lambda^{(3)}(0,0) = x_1 \Gamma
+ x_2 \gamma^{\mu_1} \gamma^{\mu_2} \Gamma \gamma_{\mu_2} \gamma_{\mu_1}
+ x_3 \gamma^{\mu_1} \gamma^{\mu_2} \gamma^{\mu_3} \gamma^{\mu_4} \Gamma
\gamma_{\mu_4} \gamma_{\mu_3} \gamma_{\mu_2} \gamma_{\mu_1}\,,
\label{Djll:struct}
\end{equation}
because the number of $\gamma$-matrices to the left of $\Gamma$ and to the right of it
has to be even (for the massless quark).
Taking separately traces of $\gamma$-matrices to the left of $\Gamma$ with $L_i$,
and of those to the right with $R_i$, where $L_i\times R_i=1\times1$,
$\gamma^{\nu_1}\gamma^{\nu_2}\times\gamma_{\nu_2}\gamma_{\nu_1}$,
$\gamma^{\nu_1}\gamma^{\nu_2}\gamma^{\nu_3}\gamma^{\nu_4}\times
\gamma_{\nu_4}\gamma_{\nu_3}\gamma_{\nu_2}\gamma_{\nu_1}$,
and solving the linear system, we can find $x_i$ via these double traces.
We may apply this procedure to the integrands of the diagrams,
thus expressing $x_i$ via scalar Feynman integrals.
Now we specialize to $\Gamma$ being the antisymmetrized product
of $n$ $\gamma$-matrices~(\ref{Djll:j}):
\begin{equation}
\gamma^{\mu_1} \gamma^{\mu_2} \Gamma \gamma_{\mu_2} \gamma_{\mu_1} = h \Gamma\,,\qquad
h = (d-2n)^2\,,
\label{Djll:h}
\end{equation}
and obtain
\begin{equation}
\Lambda_n^{(3)} = x_1 + x_2 h + x_3 h^2\,.
\end{equation}

We made minimal replacements in the code calculating the HQET vertex.
For the vector current ($n=1$), the Ward identity $\Lambda_1=-\Sigma_V(0)$
gives $\zeta_1^0=1$, to all orders of perturbation theory.
Therefore, corrections in $\zeta_n^0$~(\ref{Djll:zeta0}) are proportional to $h-(d-2)^2$.
We obtain
\begin{eqnarray}
\zeta_n^0 &=& 1 + C_F T_F \frac{g_0^4 m_{c0}^{-4\varepsilon}}{(4\pi)^d}
\left[h - (d-2)^2\right]
\Biggl[ - I_0^2 \frac{(d-2)(d-6)}{2d(d-5)(d-7)}
\nonumber\\
&&{} + \frac{g_0^2 m_{c0}^{-2\varepsilon}}{(4\pi)^{d/2}} \frac{1}{d(d-1)}
\sum_{i=1}^4 \sum_{j=1,2,5} C_i \bar{I}_j v_{ij} + \cdots \Biggr]\,,
\label{Djll:bare}
\end{eqnarray}
where all non-zero coefficients are
\begin{eqnarray}
v_{11} &=& \sigma_{11}\,,\qquad
v_{15} = \sigma_{15}\,,\qquad
v_{21} = \sigma_{21}\,,\qquad
v_{25} = \sigma_{25}\,,
\nonumber\\
v_{32} &=& \sigma_{32}\,,\qquad
v_{41} = \sigma_{41}\,,\qquad
v_{45} = \sigma_{45}\,,
\nonumber\\
v_{12} &=& - \frac{1}{2} (d-3) (d-5)^2
\Bigl[ (d-1) (d-4)^2 (d-5) (d-9) h
\nonumber\\
&&{} - 4 \left(4 d^6 - 94 d^5 + 882 d^4 - 4209 d^3 + 10681 d^2 - 13532 d + 6736\right) \Bigr]\,,
\nonumber\\
v_{22} &=& - (d-3) (d-5)
\Bigl[ (d-1) (d-4)^2 (d-5)^2 h
\nonumber\\
&&{} + 2 d^7 - 55 d^6 + 629 d^5 - 3878 d^4 + 13904 d^3 - 28928 d^2 + 32274 d - 14932 \Bigr]\,.
\label{Djll:coef}
\end{eqnarray}
The result is gauge invariant, as expected.

Now we calculate the renormalized decoupling coefficient~(\ref{Djll:zeta}):
\begin{eqnarray}
&&\zeta_n(\mu) = 1
+ \frac{1}{3} C_F T_F (n-1)
\left[ 2 (n-3) L^2 - \frac{2}{3} (n-15) L + \frac{1}{18} (85 n - 267) \right]
\left(\frac{\alpha_s(\mu)}{4\pi}\right)^2
\nonumber\\
&&{} + C_F T_F (n-1) \Biggl\{
\frac{4}{27} (n-3) \left[ 11 C_A - 4 T_F (n_l+2) \right] L^3
\nonumber\\
&&\qquad{} + \frac{2}{3} \left[ (n-3) (5 n^2 - 20 n - 8) C_F
- \frac{1}{3} (n-3) (6 n^2 - 24 n - 29) C_A
+ \frac{4}{9} (n-15) T_F \right] L^2
\nonumber\\
&&\qquad{} + \biggl[ \left( 16 (n-3) \zeta_3
+ \frac{1}{9} (17 n^3 + n^2 - 275 n + 117) \right) C_F
\nonumber\\
&&\qquad\qquad{} - \left( 16 (n-3) \zeta_3
+ \frac{1}{81} (18 n^3 + 306 n^2 - 2003 n + 405) \right) C_A
\nonumber\\
&&\qquad\qquad{} - \frac{4}{81} (163 n - 477) T_F n_l
- \frac{496}{81} (n-3) T_F \biggr] L
\nonumber\\
&&\qquad{} + \frac{1}{3} \biggl[ 16 (n-3) B_4
- \frac{4}{5} (n-3) \pi^4 + \frac{2}{3} (187 n - 513) \zeta_3
\nonumber\\
&&\qquad\qquad{} + \frac{1}{54} (1727 n^3 - 11681 n^2 + 15946 n + 12294)
\biggr] C_F
\nonumber\\
&&\qquad{} - \frac{1}{3} \biggl[ 8 (n-3) B_4
- \frac{4}{5} (n-3) \pi^4 + \frac{1}{9} (725 n - 1887) \zeta_3
\nonumber\\
&&\qquad\qquad{} + \frac{1}{243} (3087 n^3 - 21393 n^2 + 22565 n + 49881)
\biggr] C_A
\nonumber\\
&&\qquad{} - \frac{4}{27} \left[ 32 (n-3) \zeta_3
- \frac{1}{27} (1171 n - 3981) \right] T_F n_l
\nonumber\\
&&\qquad{} + \frac{2}{27} \left[ 112 (n-3) \zeta_3
- \frac{1}{27} (1841 n - 5055) \right] T_F \Biggr\}
\left(\frac{\alpha_s(\mu)}{4\pi}\right)^3 + \cdots
\label{Djll:res}
\end{eqnarray}

The vector current does not renormalize ($Z_1=1$, $Z'_1=1$),
and we have $\zeta_1(\mu)=1$ to all orders.
This is to be expected;
for example, for a diagonal flavour matrix $\tau$ the integral of the vector current
is a combination of the differences of the full numbers of quarks and antiquarks
of several flavours,
and these differences are some integers which are the same in both theories.
Decoupling of the scalar current is related to that of the mass:
\begin{equation}
\zeta_0(\mu) = \zeta_m^{-1}(\mu)\,,
\label{Djll:mass}
\end{equation}
to all orders.
Our result~(\ref{Djll:res}) at $n=0$ reproduces
the three-loop mass decoupling~\cite{CKS:98}.

The currents $j(\mu)$ with $n=4$, 3 differ from those with $n=0$, 1
by insertion of the 't~Hooft--Veltman $\gamma_5^{\mbox{\scriptsize HV}}$.
They differ from the corresponding currents
with the anticommuting $\gamma_5^{\mbox{\scriptsize AC}}$ by finite renormalizations:
\begin{eqnarray}
\left(\bar{q} \gamma_5^{\mbox{\scriptsize AC}} \tau q\right)_\mu
&=& Z_P(\alpha_s(\mu))
\left(\bar{q} \gamma_5^{\mbox{\scriptsize HV}} \tau q\right)_\mu\,,
\nonumber\\
\left(\bar{q} \gamma_5^{\mbox{\scriptsize AC}} \gamma^\alpha \tau q\right)_\mu
&=& Z_A(\alpha_s(\mu))
\left(\bar{q} \gamma_5^{\mbox{\scriptsize HV}} \gamma^\alpha \tau q\right)_\mu\,.
\label{Djll:Larin}
\end{eqnarray}
The finite renormalization constants $Z_P$, $Z_A$ are known
with three-loop accuracy~\cite{LV:91,L:93}.
Inserting $\gamma_5^{\mbox{\scriptsize AC}}$ does not change the decoupling coefficient.
Therefore,
\begin{equation}
\frac{\zeta_4(\mu)}{\zeta_0(\mu)} =
\frac{Z'_P(\alpha_s'(\mu))}{Z_P(\alpha_s(\mu))}\,,\qquad
\frac{\zeta_3(\mu)}{\zeta_1(\mu)} =
\frac{Z'_A(\alpha_s'(\mu))}{Z_A(\alpha_s(\mu))}\,.
\label{Djll:Larin2}
\end{equation}
We have checked that our result~(\ref{Djll:res}) satisfies these strong checks.

The result for the tensor current is new:
\begin{eqnarray}
&&\zeta_2(\mu) = 1
+ \frac{1}{3} C_F T_F \left( - 2 L^2 + \frac{26}{3} L - \frac{97}{18} \right)
\left(\frac{\alpha_s(\mu)}{4\pi}\right)^2
\nonumber\\
&&{} + C_F T_F \Biggl\{
- \frac{4}{27} \left[ 11 C_A - 4 T_F (n_l+2) \right] L^3
+ \frac{2}{3} \left( 28 C_F - \frac{53}{3} C_A - \frac{52}{9} T_F \right) L^2
\nonumber\\
&&\qquad{} + \frac{1}{9} \left[ - (144\zeta_3+293) C_F
+ \left( 144 \zeta_3 + \frac{2233}{9} \right) C_A
+ \frac{604}{9} T_F n_l + \frac{496}{9} T_F \right] L
\nonumber\\
&&\qquad{} - \frac{1}{3} \left( 16 B_4 - \frac{4}{5} \pi^4 + \frac{278}{3} \zeta_3
- \frac{5639}{27} \right) C_F
+ \frac{1}{3} \left( 8 B_4 - \frac{4}{5} \pi^4 + \frac{437}{9} \zeta_3
- \frac{34135}{243} \right) C_A
\nonumber\\
&&\qquad{} + \frac{4}{27} \left( 32 \zeta_3 - \frac{1639}{27} \right) T_F n_l
- \frac{2}{27} \left( 112 \zeta_3 - \frac{1373}{27} \right) T_F \Biggr\}
\left(\frac{\alpha_s(\mu)}{4\pi}\right)^3 + \cdots
\label{Djll:tensor}
\end{eqnarray}

\end{document}